# Direct image of surface plasmon-coupled emission by leaky radiation microscopy

D.G. Zhang [1], X.-C.Yuan [2*], A. Bouhelier [3]

1School of Electrical and Electronic Engineering, Nanyang Technological University, Nanyang Avenue, Singapore 639798

2Institute of Modern Optics, Key Laboratory of Optoelectronic Information Science & Technology, Ministry of Education of China, Nankai University, Tianjin 300071, People's Republic of China

3Institut Carnot de Bourgogne, CNRS-UMR 5209, Université de Bourgogne, 21078 Dijon, France

*Corresponding author: xcyuan@nankai.edu.cn

Leaky radiation microscopy (LRM) is used to directly image the surface plasmon-coupled emission (SPCE). When compared with the prism based set-up commonly used in SPCE research, LRM has the advantages of directly giving out the emitting angle without scanning and the image of generated surface plasmons polaritons' (SPPs) propagation, which help to understand the optical process of SPCE. LRM also can give out clearer SPCE image than that by prism-based set-up. Based on the LRM, we find that the SPCE pattern and propagation of SPPs can be modified by the shape of samples fabricated on the thin metallic films.

## 1. Introduction

Fluorescence detection is one of the basic measurements in the biological sciences, biotechnology, and medical diagnostics. While fluorescence emission is nearly isotropic in space due to the isotropic distribution of fluorescence, it is difficult to capture more than small fraction of the total emission[1]. In the past years, surface plasmon-coupled emission (SPCE) was proposed to make the fluorescence emit at a particular angle just like a beam. The SPCE depends on localization of the samples near a thin metallic (always silver or gold) film on a transparent substrate, typically 10–200nm from the metallic surface. At these distances the emission couples with the surface plasmon polaritions (SPPs) of the metallic film and enters the transparent substrate at the surface plasmons resonance (SPR) angle. This coupling can be highly efficient to over 90% for molecules with the proper orientation and distance from the metallic surface[2, 3, 4]. Prism based set-up was commonly used to investigate the SPCE phenomenon[5, 6, 7, 8, 9, 10]. For this kind setup, the fluorescence emission angle can be measured by scanning the rotary stage, but the fluorescence conversion to SPPs and the behavior of generated SPPs on metallic film can not be imaged. In this letter, leaky radiation microscopy (LRM) is used to characterize the SPCE. LRM is a very useful tool in SPR technique[11, 12, 13, 14, 15]. In SPR techniques where the excitation area roughly compares with the SPPs propagation length, this lossy leakage radiation (LR) interferes destructively with the incoming excitation light and can not be detected. However, if the excitation area is significantly smaller than the lateral spread of the SPPs (using high N.A objective to focus the exciting beams), LR can be observed. The observation of LR gives a direct measurement of the non-radiative SPPs traveling at the opposite interface. The intensity of the losses, at a given lateral position in the film, is proportional to that of the SPPs at the same position. This is the working principle of the LRM[16, 17, 18]. So when LRM is used to characterize

SPCE, both the emitting angle and propagation of excited SPPs wave can be directly imaged. In this letter, LRM is used to characterize SPCE from Rhodamine (RhB) doped PMMA stripe or PMMA/Air boundary fabricated on silver film. Experimental results show that the SPCE pattern can be modified by the shape of the sample although the doped RhB are of random orientation.

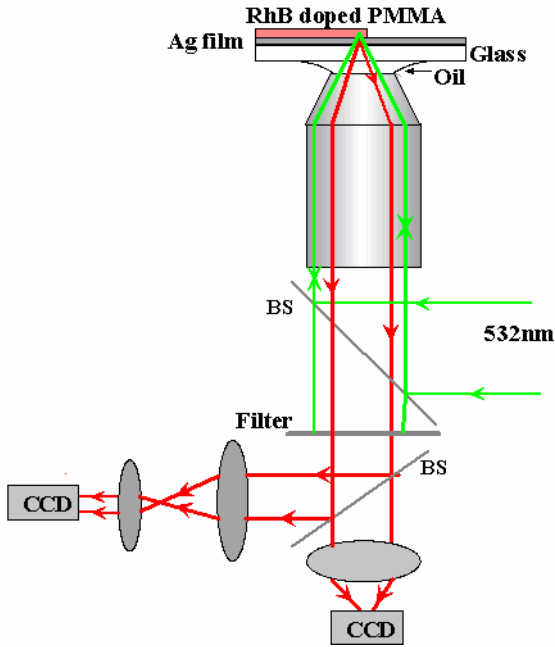

Fig.1. Sketch of the experimental set-up. BS is beam splitter.

2. **Experiments**

The samples were prepared as follow. Rhodamine (RhB) molecules (0.1mg/ml) were dissolved in a PMMA solution (950K PMMA, Solids: 2% in Anisole, from MICRO.CHEM corp.) for about 48 hours. The doped solution was agitated by ultrasonic disrupter for 30 minutes before the spin-coating process. The RhB doped PMMA film used in this experiment was obtained by spin coating the solution onto 45 nm-thick silver films deposited by electron beam evaporation on a glass substrate. The film was baked for 10min at 105$^\circ$C to remove the solvent.

Electron beam lithography (Raith GmbH, e_LiNe) and standard developing procedure were used to write PMMA stripes or boundary between PMMA and air on the Ag film.

A 532 nm wavelength laser was used to excite the RhB molecules by using an oil-immersion objective (60X, numerical aperture (N.A.), 1.42). The laser beam was expanded with a lens assembly to overfill the back aperture of the objective. Excitation of the RhB and detection of the fluorescence signal was collected by the same objective (Epi-configuration). The power of the 532nm laser before entering the objective was about 0.05mW. A 532 nm long pass edge filter was placed before a charge-coupled device (CCD) camera to reject the exciting beam. Both the Fourier plane image and direct image plane were captured by the CCD cameras as shown in Figure 1. The thickness of the PMMA film is about 80nm.

3. **Results and discussions**

Figure 2 (a) shows the direct fluorescence image with the exciting beam focused onto the boundary between the PMMA and Air. The white dash line represents the boundary between the PMMA film and air. The half bright spot in the PMMA side displays that RhB molecules doped in the PMMA film were excited and emit fluorescence. On the air side, there is wave propagating always from the boundary and attenuating during propagation. If we move the laser focus totally on the air side, no such wave appears. We attribute this as the SPPs generated by the excited RhB molecules at the boundary of PMMA film. To verify this point, the Fourier plane fluorescence image of the microscope was captured and shown in Figure 2 (b). It shows that there one bright full ring at the edge and a right-hand side ring inside. For comparison, the Fourier plane fluorescence image with the exciting beam totally focused onto the PMMA side was captured and shown in Figure 2(c), which displays only one bright ring at the edge. The two kind rings are the common known SPCE, which are of different SPPs existing on the silver film. The larger ring is related with SPPs generated on the Air/PMMA/Ag interface, and the inside half ring is due to the SPPs on the Air/Ag interface. It should be noted that the SPCE ring represents the wave-vector (value and propagating direction) of the SPPs generated by the excited fluorescence molecules.

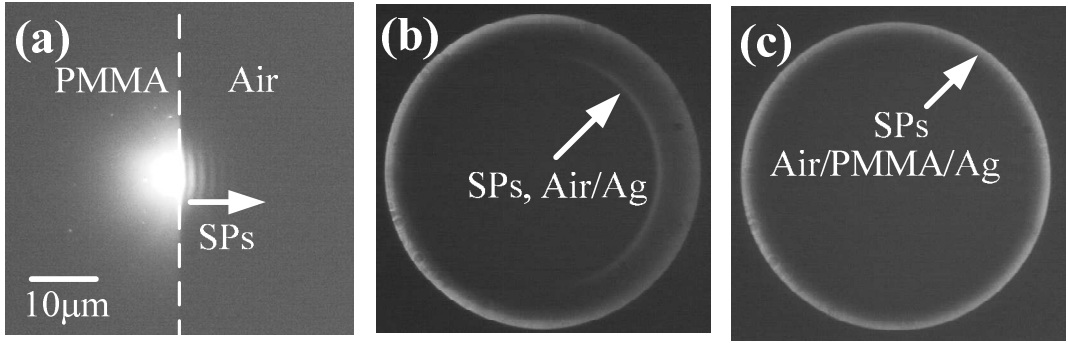

Fig.2. Excitation of SPPs (Air/Ag interface and Air/PMMA/Ag) with RB doped in PMMA films (a): Florescence image; (b) and (c): Fourier plane fluorescence image with exciting beam focused on the boundary and PMMA part. The white dash line represents the boundary between Air and PMMA.

Figure 2 (c) shows the full ring with uniform field intensity on every point of the ring which represents the generated SPPs propagating in all directions on the Ag film due to the random orientation of the molecules insider the extended PMMA film. While Figure 2 (b) shows that the intensity of the left-hand side on the larger ring is stronger than that of right-hand side. It is because SPPs (Air/PMMA/Ag) mainly exist in the left side (PMMA part). When it propagates into the right hand side (air part), it diminishes. So the intensity on the right-hand side ring caused by the SPPs (Air/PMMA/Ag) propagating to the right side is weaker. Because the SPPs (Air/Ag) are generated by the excited RhB doped in PMMA film and do not exist in the Air/PMMA/Ag multilayer, it will only propagate into the right side (air side), so only half ring appears. The intensity of half ring is strong in the center and diminishes away from the center, which represent that the SPPs (Air/Ag) mainly propagate in the direction vertical to the boundary, and is consistent with the LR image of SPPs wave shown in Figure 2(a). These phenomena give out an interesting conclusion: Although the fluorescence molecules are of random orientation in the PMMA film, the propagating direction of the SPPs (Air/Ag) on the silver film generated by excited fluorescence molecules is not uniform in all the directions, but selective, which means the shape of the sample influences the generated SPPs' propagating direction or the intensity distribution of SPCE ring (SPCE pattern).

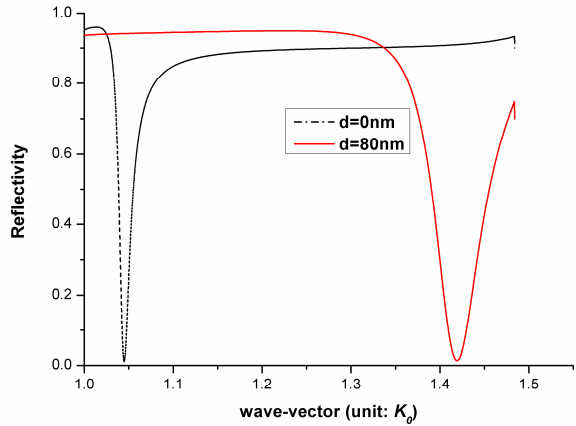

Fig.3: SPR curves with PMMA film thickness d=0nm and d=80nm for two kind SPPs wave generated on Air/Ag interface and Air/PMMA/Ag interface.

Based on the known N.A. of the objective, the wave-number of full and half ring can be estimated as 1.419 $K_0$ and 1.049 $K_0$ ($K_0$ is the vacuum wave-vector of the RhB fluorescence peak taken at 576 nm). Numerical calculation were carried out to get the *Ksp* of the SPPs exist in the Air/PMMA/Ag or Air/Ag interface. The SPR curves are shown in Fig. 3, which shows the SPR angle increases with PMMA film on silver film. In the calculation, the refractive indexes of the PMMA, Ag and glass substrate are 1.49, 0.12+3.547i, 1.516 respectively[19]. The wavelength is 576nm which is the fluorescence peak of RhB doped in PMMA film. The calculated *Ksp* of SPPs are 1.0449 and 1.4194 respectively, which are consistent with the value calibrated from Figure.2 respectively and further verify that the two rings are attributed to the two kind SPPs.

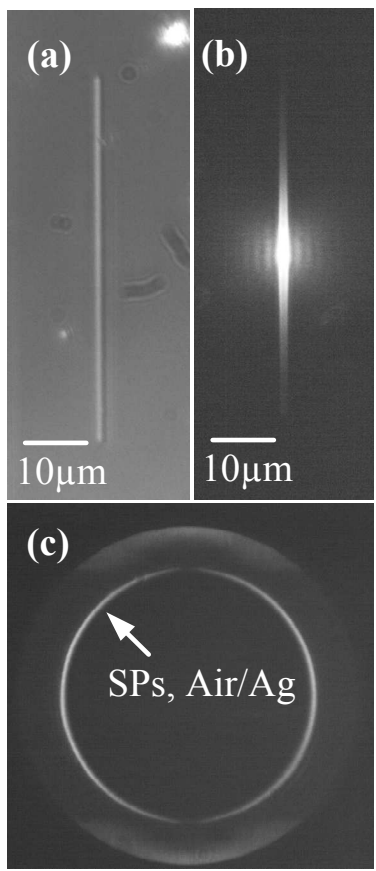

Fig.4. (a) Bright-field transmission image of PMMA stripe; (b) is direct-space fluorescence image; (c) Fourier plane fluorescence image.

To further investigate the influence of shape on the SPCE pattern, we have employed another sample shape. Here we fabricate one RhB doped PMMA stripe with electron beam lithography (EBL) on the silver film as shown in Figure 4(a). The thickness, width and length of the stripe are 80nm, 1µm and 60µm respectively. The power of the pump beam was about 0.15mw. Figure 4(b) is the fluorescence image with the exciting beam focused onto the center of the stripe. It shows that the intensity of fluorescence is strong in the central and attenuate along the stripe. On both side of the stripe, there are two waves propagating on the silver film vertically to the stripe and attenuating away from the stripe, which are the leaky radiating SPPs wave (Air/Ag). Fourier plane fluorescence image was captured in Figure 4(c) which two half bright rings appear on right-hand and left-hand side with two gaps (up and down) between them. These two half rings are attributed to the SPPs (Air/Ag) on both side of the stripe. The wave-number of the SPPs wave corresponding to the SPCE ring can be estimated as 1.052 K0, and is approximate to the calculated value shown in Figure 3. The bright fluorescence lines on the upside and bottom of Figure 4(c) are attributed to the modes existing in the PMMA stripe, such as $TM_{00}$ (SPPs, Air/PMMA stripe /Ag), $TM_{01}$, $TM_{02}$ modes and so on. It is difficult to separate these modes in the Fourier plane image, because of the spectrum width of the fluorescence. The intensity of the two SPCE rings is strong at the center and become weak along the ring, which is consistent with that shown in Figure 2(b) and also consistent with Figure 4(b) displaying the propagation of the generated SPPs wave. The difference between Figure 4(c) and Figure 2 (b) is that there are two SPPs (Air/Ag) waves generated in Figure 4(c) and inducing two half rings with similar field intensity distribution. So Figure 2 and Figure 4 both verify that the SPCE ring's intensity distribution (SPCE pattern) or the propagation of the SPPs can be modified by the samples' shape.

## 4. Summary

In summary, LRM is used to directly image the SPCE, which directly gives out both the emitting angle without scanning and the image of SPPs' propagation simultaneously. The observation of SPCE with LRM gives out the clear physical pictures of the process of SPCE: The exciting fluorescence molecules transfer to SPPs propagating on the thin metallic film. During the propagation, the SPPs leaky radiate into the glass substrate at the SPR angle, which forms the commonly known SPCE. The field intensity distribution of the SPCE ring is related with the propagating directions of the SPPs on the metallic film which can also been imaged by the LRM. Figure 2 and Figure 4 also show that the SPCE ring images obtained by LRM are clearer than that by prism-based set-up. It is because the back focal plane image of the objective is the Fourier transform of the image plane which contains all the information in the K-space[20].

Based on the LRM set-up, we find that SPCE pattern can be influenced by the shape of the samples. If the fluorescence molecules (randomly oriented) under excitation are doped in isotropic structures, such as extended PMMA film as in Figure 2, the excited molecules will transfer to SPPs with isotropic propagating directions, which induce the entire SPCE ring shown in Figure 2(c). On the contrary, if the molecules (randomly oriented) are doped in anisotropic structures, such as the PMMA/Air boundary or PMMA stripe, the intensity distribution of the SPCE (SPCE pattern, or the propagating direction of the generated SPPs) will be adjusted as shown in Figure 2(b) and Figure 4(c). In Ref 9 and 21, the authors also demonstrate a method to modify the SPCE pattern or propagation of SPPs by the oriented 6T layer (orientation of fluorescence molecules), which is different from the modification of SPCE pattern shown here by the shape of samples containing the randomly oriented fluorescence molecules. So the two experimental results show that the molecules orientation and samples' shape both can modify the SPCE pattern and SPPs' propagation, which have potential applications in radiative decay engineering[2,3] or plasmonics devices as proposed in the recent theoretical papers[22,23]. The future work involves various shape of the samples fabricated on silver film to realize different SPCE patterns or control SPPs' propagation.

This work was partially supported by the National Natural Science Foundation of China under Grant No. (60778045), the National Research Foundation of Singapore under Grant No. NRF-G-CRP 2007-01 and the Ministry of Science and Technology of China under Grant no. 2009DFA52300 for China-Singapore collaborations. XCY acknowledges the support given by Nankai University (China) and Nanyang Technological University (Singapore).